\documentclass[10pt,twocolumn]{article}

\usepackage[margin=1.8cm,top=2cm,bottom=2cm]{geometry}
\usepackage{amsmath,amssymb,amsthm}
\usepackage{graphicx}
\usepackage[table,xcdraw]{xcolor}
\usepackage{booktabs}
\usepackage{tabularx}
\usepackage{microtype}
\usepackage{caption}
\usepackage{subcaption}
\usepackage{float}
\usepackage{enumitem}
\usepackage{cite}
\usepackage{url}
\usepackage{balance}
\usepackage{dblfloatfix}
\usepackage{cuted}
\usepackage{capt-of}

\usepackage{XCharter}
\usepackage[charter]{mathdesign}
\usepackage{titlesec}
\titleformat{\subsection}{\small\bfseries}{\thesubsection}{1em}{}
\newcommand{\Jt}{J^{\top}}
\newcommand{\rhoone}{\rho_1}
\newcommand{\alphac}{\alpha_c}
\newcommand{\xctilde}{\tilde{x}_c}
\newcommand{\Deltax}{\Delta\tilde{x}_c}
\newcommand{\dmg}{\delta_{\mathrm{mg}}}
\newcommand{\Fvis}{F_{\mathrm{visible}}}
\newcommand{\Vsoft}{V_{\mathrm{soft}}}
\newcommand{\Vstiff}{V_{\mathrm{stiff}}}

\begin{document}

\title{\textbf{Latent Geometry as a Structural Monitor:}\\
Eigenspace Alignment for Anomaly Detection\\
in Anonymity Networks}

\author{Vaibhav Chhabra\\
{\small USPTO Patent \#64/034,500}}

\date{}
\maketitle

\begin{abstract}
Traditional anomaly detection marks events when measured
signals cross predefined thresholds. This captures the
moment of transition but not the structural pressure that
precedes it. We propose treating large behavioral
populations as geometric energy landscapes whose deformation
can be measured before and during major transitions. The
central thesis is that \emph{structure precedes geometry}:
the structural organization of the population is the signal,
and geometric metrics are instruments for measuring it.
Applied to the Tor anonymity network across 67 consecutive
daily observation windows, the dual-observer pipeline
identifies a stable nine-dimensional load-bearing subspace
invariant across the observation period and validates this
structure by Monte Carlo simulation at $16.8\sigma$ above
the noise floor. Primary detection gates achieve 0.0\% false
positive rate~(FPR) on 24 confirmed stable windows. Forensic
analysis of the February 20, 2026 confirmed infrastructure
event formally falsifies the relay-departure hypothesis,
identifying connectivity degradation without topology change
as a detectable network failure mode. The result is a
candidate structural-monitoring framework for behavioral
populations with sufficient telemetry.
\end{abstract}

\section{Introduction}

Network security monitoring has historically relied on a
simple premise: define what an attack looks like, then watch
for it. Signature-based intrusion detection systems,
threshold-based alerting, and rule-based anomaly detectors
share this structure. They are reactive by design, optimized
to recognize known patterns after they have already
manifested.

This approach has a structural limitation. It detects events,
but not the conditions that produce them. A network under
sustained structural pressure can appear healthy until a
threshold is crossed. At that moment the alert fires, but the
pre-transition deformation has already accumulated.

We propose a different framing grounded in the principle
that structure precedes geometry. A large behavioral
population, such as a network of relays, can be modeled as
occupying a learned geometric manifold. The encoder learns
which directions in feature space are structurally
load-bearing (the \emph{stiff axes}) and which are elastic
(the \emph{soft axes}). Under normal operation, population
movement is absorbed through the soft axes. Structural
events force movement into the stiff axes, producing a
stiff-axis collision that signals stress. The formal
definition of stiff and soft axes is given in
Section~\ref{sec:ejt}.

This work is best understood as a \emph{systems-observation
framework paper}. It does not claim attack attribution,
prediction certainty, or packet-level causality. The Tor
anonymity network withholds those forms of visibility by
design. The contribution is to show that structural
deformation of behavioral populations can be measured before
and during major transitions using only public relay
metadata.

Throughout the paper, thermodynamic and physical terminology
(stiff, elastic, fracture, mass, gravity) is used
analogically. Each term is operationalized mathematically
before use. The analogies are instruments of description,
not claims about physical mechanism.

\textbf{Contributions.} We introduce a dual-observer
pipeline combining a Contractive Denoising Autoencoder
(CDAE)~\cite{vincent2010,rifai2011} as a geometric observer and a Gaussian Restricted
Boltzmann Machine (GRBM)~\cite{hinton2002} as a thermodynamic observer,
bridged by Canonical Correlation Analysis~(CCA). We show
that the Eigendecomposition of the Jacobian Trace~(EJT)
produces a stable nine-dimensional stiff subspace invariant
across 67 consecutive observation windows. We formally
falsify the relay-departure hypothesis for the February 20,
2026 confirmed infrastructure event and identify
connectivity degradation without topology change as a
previously invisible failure mode. Finally, we identify a
coordinated residential relay deployment coinciding with the
February 05--13 population surge, while leaving operator
intent unresolved.

\section{Related Work}

This work sits at the intersection of Tor network monitoring,
network anomaly detection, manifold learning, Jacobian-based
sensitivity analysis, and multi-view representation learning.
The relevant prior work is mature in each individual area, but
less developed at their intersection: structural monitoring of
an encrypted behavioral population through agreement between
independent latent observers.

\textbf{Tor network monitoring.} The Tor Project's public
metrics, Onionoo metadata, and directory-consensus data~\cite{torproject,onionoo} enable
relay-level analysis without exposing packet contents or user
traffic. Prior Tor security work has focused on traffic
analysis, relay deanonymization, consensus manipulation,
Sybil detection, relay fingerprinting, bandwidth anomalies,
and autonomous-system concentration. This paper differs by
operating on behavioral population geometry rather than
individual relay identity. It uses public relay metadata and
requires no labeled attack data.

\textbf{Network anomaly detection.} Unsupervised anomaly
detection in network telemetry commonly uses statistical
process control, autoencoders, clustering, or one-class
classification. Reconstruction-error autoencoders detect
that distributional drift occurred, but they do not explain
which structural directions absorbed the change. Here, the
encoder Jacobian is decomposed into stiff and soft subspaces,
so the output is not only an anomaly signal but a geometric
classification of the event type.

\textbf{Manifold learning and behavioral geometry.} Methods
such as PCA, UMAP, and t-SNE~\cite{umap,tsne} are widely used to visualize
behavioral data and reveal clusters. These embeddings are
usually static. The contractive autoencoder used here learns
a representation whose local sensitivity structure can be
tracked across time, turning manifold learning from a display
method into a monitor of structural deformation.

\textbf{Jacobian sensitivity.} Jacobian analysis in neural
networks has been used for robustness evaluation,
adversarial example detection, and feature importance. The
metric tensor approximation $\Jt J$ used here is related to
local sensitivity and Fisher-information-style geometry. The
novel use in this paper is to interpret the eigenspectrum of
$\Jt J$ as a load-bearing structure for anomaly detection in
network behavior.

\textbf{Multi-view representation learning.} Canonical
Correlation Analysis~\cite{hotelling1936} has long been used to identify shared
structure between heterogeneous views. We use CCA to bridge
CDAE and GRBM latent representations and monitor the
magnitude and rotation of observer agreement over time. This
turns multi-view alignment into a temporal structural signal.

\section{Background}

\textbf{Tor relay metadata.} Tor is a volunteer-operated
anonymity network of approximately 7,000--11,000 active
relays. Relays are assigned roles by the directory authority
consensus. \emph{Guard} relays serve as entry points for
client circuits. \emph{Exit} relays allow traffic to leave
Tor into the open internet and are rarest, comprising about
20\% of the network. \emph{Middle} relays route traffic
internally. The Onionoo API~\cite{onionoo} provides historical metadata
derived from hourly consensuses. Tor provides no packet-level
visibility, so the framework operates entirely on relay
behavioral metadata.

\textbf{Observation windows.} Each observation window is one
consecutive day-pair: the Tor relay population snapshot at
date $t$ paired with the snapshot at date $t+1$. Detection
signals are computed as deltas between consecutive pairs. We
observe 67 consecutive windows spanning January~19 to
April~21, 2026. The EJT z-score baseline was established
from 14 stable windows (January~13--26, 2026), yielding
frozen constants mean $=0.750$ and standard deviation
$=0.113$. The per-cluster EJT uses the
consensus-weight-weighted median as the cluster evaluation
point; the global EJT z-score uses the arithmetic population
mean. These are distinct computations producing distinct
signals.

\textbf{Feature space and null baseline.} Relay behavior is
captured by 191 Onionoo-derived features. To establish a
noise floor, we ask whether the pipeline signal could be an
architectural artifact: if so, random Gaussian noise passed
through the same trained pipeline would produce similar
outputs. Over 1,000 Monte Carlo iterations, the null baseline
yields $\rhoone=0.509$ with maximum 0.614. Empirical Tor data
yields $\rhoone=0.9992$. The $16.8\sigma$ separation confirms
that the measured structural agreement is a property of the
Tor relay population rather than the pipeline alone.

\textbf{Failure modes.} Three failure modes are relevant.
First, \emph{topology change}: relays depart or join the
consensus, which standard relay-count monitoring can detect.
Second, \emph{behavioral shift}: relays remain present but
change observable features, which distributional methods may
detect. Third, \emph{connectivity degradation without
membership change}: routing paths are severed while relays
remain listed in the consensus, which topology monitoring
misses. The February 20, 2026 confirmed event provides
evidence of the third type.

\section{Architecture}

\subsection{Overview}

The framework relies on a dual-observer pipeline. A CDAE
functions as the geometric observer, mapping the structural
geometry of the relay population manifold. A GRBM functions
as the thermodynamic observer, evaluating the population's
energy landscape. CCA serves as the mathematical bridge
between these observers. The EJT defines the load-bearing
axes. These components were not designed with a prescribed
firing order. Empirically, in the January 23--26 to January
27 sequence, the GRBM thermodynamic signal (Ch5~CV) preceded the
CDAE geometric signal. Whether this sequentiality holds
across all event types is addressed in
Section~\ref{sec:signal_rel}.

\begin{figure*}[t]
  \centering
  \includegraphics[height=0.78\textheight,keepaspectratio]{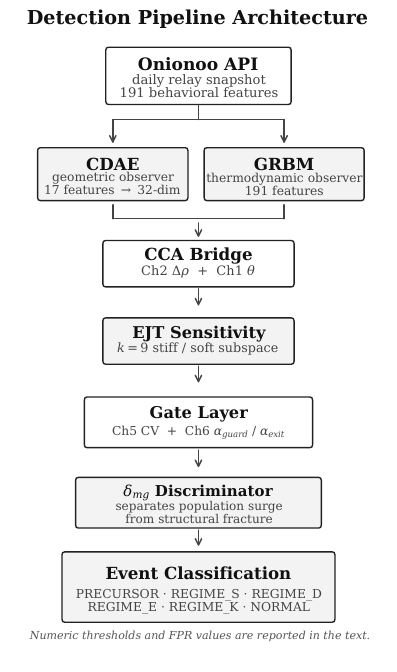}
  \caption{Pipeline architecture. Daily Onionoo relay snapshots enter
    the CDAE geometric observer and GRBM thermodynamic observer. CCA
    bridges observer agreement, EJT defines stiff and soft sensitivity
    directions, gate layers classify structural events, and $\dmg$
    separates population surge from structural fracture.}
  \label{fig:architecture}
\end{figure*}

Figure~\ref{fig:architecture} illustrates the gate cascade
from daily relay population input through all six channels
to event classification.

\subsection{Input Space}
The raw input space consists of 191 behavioral features
from the Tor Onionoo API~\cite{onionoo}. For geometric sensitivity
analysis, 17 clean continuous features are selected
(CLEAN\_INDICES $= [1,3,4,5,6,7,8,9,10,\linebreak 11,13,14,28,29,
30,31,32]$). Binary flags are excluded because discrete
variables produce infinite or zero Jacobian jumps at
transition points. These 17 features span four behavioral
categories. \emph{Capacity}: observed bandwidth, bandwidth
rate, bandwidth burst, advertised bandwidth, bandwidth
difference, bandwidth ratio, and burst-to-rate ratio.
\emph{Role assignment}: middle probability, guard
probability, exit probability, consensus weight, and
consensus weight fraction. \emph{Geographic and temporal
stability}: latitude, longitude, relay lifespan in days,
and days since last restart. \emph{Health}: overload
general timestamp. The continuous features exhibit moderate
right-skew (mean/median ratio $= 1.4$) with isolated
extreme outliers (max/median $= 14\times$). All cluster
centers are computed as the consensus-weight-weighted
median, anchoring evaluation to the relay most
representative of actual Tor traffic load. This choice
reduced exit cluster FPR from 26.3\% to 0.0\%, validated
in Section~\ref{sec:pop_center}.

\subsection{CDAE: Geometric Observer}
The geometric observer is a deterministic autoencoder.
While initially designed as a Variational Autoencoder
(VAE), the variance head produced NaN instabilities during
training due to isolated extreme outliers, and was retired.
The model maps 17 scaled continuous features to a
32-dimensional latent space. The 32-dimensional latent
space provides an overcomplete representation
(approximately $2\times$ the 17 input features), giving
sufficient degrees of freedom for a meaningful stiff and
soft subspace decomposition. The training loss function is:
\begin{equation}
  \mathcal{L} = \|x - \hat{x}\|^2
              + \lambda_c \|J\|_F^2,
  \quad \lambda_c = 0.001
\end{equation}
The Frobenius norm $\|J\|_F^2$ penalizes large Jacobian
entries, forcing the encoder to suppress sensitivity to
noisy high-variance directions. Encoder weights are frozen
after the training window (January~19 -- February~09, 2026).

\subsection{GRBM: Thermodynamic Observer}
The thermodynamic observer is a GRBM trained on the full
191-feature space. \emph{Sigmoid saturation} occurs when
input values are so large that the sigmoid activation
$\sigma(z) = 1/(1+e^{-z})$ outputs values near 0 or 1 for
all units, causing gradients to vanish and the model to
stop learning. Tor's extreme bandwidth outliers ($14\times$
the median) caused sigmoid saturation in the hidden layer
at inference time, rendering hidden activations unusable.

The solution: instead of reading the hidden layer output
(which passes through the sigmoid), we read the visible
free energy directly from the visible layer, before the
sigmoid is applied:
\begin{equation}
  \Fvis(x) = \tfrac{1}{2}\sum_i
  \frac{(x_i - b_i)^2}{\sigma_i^2}
\end{equation}
This is the squared deviation of each relay's features
from the learned population baseline $(b_i, \sigma_i)$.
The computation never passes through the sigmoid, so
saturation cannot occur. The trained parameters were
verified well-conditioned (hidden bias mean $= -0.019$,
std $= 0.003$, no unit with $|b_h| > 5$), confirming the
model learned correctly and saturation was an
inference-time distributional mismatch, not a training
failure. The thermodynamic fragmentation signal is the
Coefficient of Variation (CV) of $\Fvis$ across the
relay population per window:
$\mathrm{CV} = \mathrm{std}(\Fvis) / \mathrm{mean}(\Fvis)$.
A high CV indicates the population is fragmenting into
distinct behavioral groups.

\subsection{CCA Bridge}
Canonical Correlation Analysis (CCA) finds linear
combinations of two sets of variables that are maximally
correlated. Given the CDAE latent representation $z_{\rm
cdae} \in \mathbb{R}^{32}$ and the GRBM hidden activations
$z_{\rm rbm} \in \mathbb{R}^{32}$, CCA finds directions
in each space where the two observers most agree. The first
canonical correlation $\rhoone$ measures agreement strength;
the angle $\theta$ between consecutive first canonical
directions measures how much shared structure rotated from
one window to the next; $\Delta\rho$ measures whether
agreement strengthened or weakened.

CCA is refitted per observation window. An initial frozen
CCA fit \emph{degenerated} to $\rhoone = 0.000$ within days
as the relay population surged from ${\sim}10{,}500$ to
${\sim}19{,}640$ relays. \emph{CCA degeneration} occurs
when the two input spaces share no meaningful variance
structure, collapsing all canonical correlations to zero.
Per-window refit was adopted as the operational standard.
CCA identifies linear shared structure only; this
limitation is discussed in Section~\ref{sec:limitations}.

\subsection{EJT: Eigendecomposition of Jacobian Trace}
\label{sec:ejt}
The Jacobian $J = dz/dx$ evaluated at a cluster center
$\xctilde$ is a matrix of partial derivatives. The metric
tensor approximation:
\begin{equation}
  \Jt J = J_{\rm clean}^\top J_{\rm clean},
  \quad \text{shape } (17 \times 17)
\end{equation}
captures the combined sensitivity of all latent dimensions
to each pair of input features. Eigendecomposition of
$\Jt J$ produces eigenvectors ordered by descending
eigenvalues. The top $k = 9$ eigenvectors define the
\emph{stiff subspace} $\Vstiff$: directions the encoder
is most sensitive to, where it resists deformation. The
remaining 8 define the \emph{soft subspace} $\Vsoft$:
directions the encoder barely notices, where the population
can move freely.

The threshold $k = 9$ was not chosen; it emerged as the
only value invariant across all 67 windows at the 90\%
trace mass criterion, validated in
Section~\ref{sec:stiff_stability}.

The EJT is evaluated separately for each role cluster
(Guard, Middle, Exit) using the consensus-weight-weighted
median as evaluation point, and also at the global
population arithmetic mean. These produce different
z-scores and may disagree. When the global EJT fires but
per-cluster EJT remains quiet, the event affected the
population as a whole without concentrating in any
single role cluster.

The EJT soft alignment ratio $\alphac$ for cluster $c$:
\begin{equation}
  \alphac = \frac{\|\Vsoft\Vsoft^\top\Deltax\|^2}
                 {\|\Deltax\|^2}
\end{equation}
When $\alphac \approx 1$, the cluster moved through soft
directions: elastic absorption. When $\alphac \approx 0$,
the cluster moved into stiff axes: a stiff-axis collision
signaling structural stress.

\subsection{Mass-Gravity Divergence ($\dmg$)}
The mass-gravity divergence $\dmg$ distinguishes two events
that can produce identical $\alpha$ signals: a genuine
structural fracture and a benign population surge of
lightweight relays.

\emph{In data science terms}: the consensus-weight-weighted
median (gravity) is anchored to where Tor traffic flows.
The unweighted median (mass) counts every relay equally.
When lightweight relays flood in, mass shifts while
gravity barely moves.

\begin{figure*}[t]
  \centering
  \includegraphics[width=\textwidth]{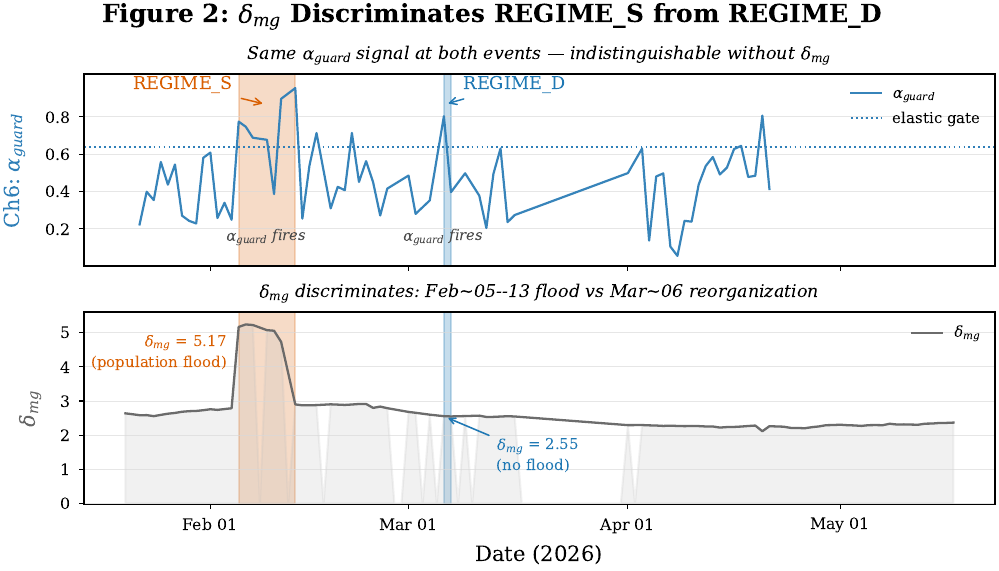}
  \caption{$\dmg$ discriminator. The Feb~05--13 surge and Mar~06
    reorganization show similar $\alpha_{\rm guard}$ activity but
    different $\dmg$ values, separating population-mass movement
    from internal cluster reorientation.}
  \label{fig:dmg}
\end{figure*}

\emph{In physics terms}: gravity is pulled toward heavy,
load-bearing relays. Mass gives equal weight to every relay.
Divergence between mass and gravity signals a lightweight
relay influx without traffic-load contribution.
\begin{equation}
  \dmg = \|\tilde{x}_{\rm weighted} - \tilde{x}_{\rm median}\|
\end{equation}
A large $\dmg$ with high $\alpha_{\rm guard}$ indicates
an elastic population-surge signature. A lower $\dmg$ with
global stiff-axis EJT firing indicates structural stress
without a population flood. Figure~\ref{fig:dmg} illustrates
this distinction between the Feb~05--13 surge and the
Feb~20 confirmed event.

\section{Signal Validation}

\subsection{Noise Floor Establishment}
Over 1,000 Monte Carlo iterations of Gaussian noise through
the pipeline, the null baseline yields $\rhoone = 0.509$
(maximum $0.614$). Empirical Tor data yields
$\rhoone = 0.9992$. This $16.8\sigma$ separation confirms
$\rhoone \approx 0.999$ is a genuine structural property
of the Tor relay population, not an architectural artifact.

\subsection{Stiff Subspace Stability}
\label{sec:stiff_stability}
Across all 67 observation windows, the EJT requires exactly
$k = 9$ eigenvectors to capture ${\geq}90\%$ of the trace
mass of $\Jt J$. Lowering the threshold to 85\% yields
$k = 8$ invariance for only 60 of 67 windows; $k = 10$
introduces variance. $k = 9$ is the sole stable threshold,
not chosen but emergent from the network manifold.
Feature loadings are temporally consistent. The Top-10 Jaccard
similarity measures the overlap between the ten most-loaded
features in the stiff subspace of one window and the next.
A score of 1.0 means the same ten features dominate in both
windows; a score of 0.0 means no overlap at all. A median
of $0.90$ across all consecutive window pairs means that,
characteristically, nine of the ten most load-bearing features
are identical from one window to the next, confirming that
the stiff subspace loads on stable, recurring features
throughout the observation period (min $0.70$, max $1.00$).
Across all 67 observation windows, the same feature
categories consistently dominate the stiff subspace:
geographic position (latitude, longitude) and temporal
stability (\texttt{days\_since\_restart}). This was
confirmed by a feature loading analysis that evaluated
encoder sensitivity at each role cluster center (Guard,
Middle, and Exit) and identified which of the 17 input
features contributed most to the top stiff eigenvectors
across consecutive windows. Geographic position and relay
operational age represent the load-bearing behavioral
identity of trusted Tor infrastructure: the features that
distinguish a committed, geographically stable relay from
a newly arriving or unstable one.

\subsection{Gate Validation}
Primary detection gates were validated against 24 confirmed
stable windows to establish strict FPR. The FPR is the
fraction of windows where a gate fires when no real
structural event occurred.
\begin{itemize}[leftmargin=*,itemsep=1pt]
  \item \textbf{Ch5 CV gate}: CV $> 3.0$. FPR $= 0.0\%$.
  \item \textbf{Ch6 global EJT gate}: z-score $< -2.0$.
    FPR $= 0.0\%$.
  \item \textbf{Ch6 guard elastic gate}: $\alpha_{\rm guard}$
    z-score $> +2.0$ AND cluster center shift magnitude
    exceeds the population median. The shift magnitude
    $\|\Delta\tilde{x}_c\|$ is the Euclidean distance the
    cluster center moved between consecutive windows.
    Requiring this to exceed the population median filters
    out windows where $\alpha$ fired on near-zero movement,
    where directional classification is undefined.
    FPR $= 14.3\%$, reducing to $3.6\%$ when Feb 05--13
    surge windows are treated as investigated structural events
    rather than stable background.
  \item \textbf{Ch6 stiff fracture gate}: $\alpha_{\rm guard}$
    z-score $< -2.0$ AND cluster center shift magnitude
    greatly exceeds the population median, indicating a
    large coordinated movement into the stiff axes rather
    than incidental noise. This gate is retained as theoretical
    REGIME\_K logic and used only for the administrative
    maintenance forensic checklist.
  \item \textbf{Ch1 CCA rotation gate}: $\theta > 60^\circ$.
    FPR $= 49\%$. Invalid as standalone alarm; supporting
    signal only.
\end{itemize}

Gate activations for all 67 observation windows are cataloged
in Table~\ref{tab:activations}. The temporal signal patterns
underlying each gate are illustrated in Figure~\ref{fig:seq}
(selected channels) and the full forensic monitor (Appendix~A).

\subsection{Signal Relationships}
\label{sec:signal_rel}
\begin{figure*}[t]
  \centering
  \includegraphics[width=\textwidth]{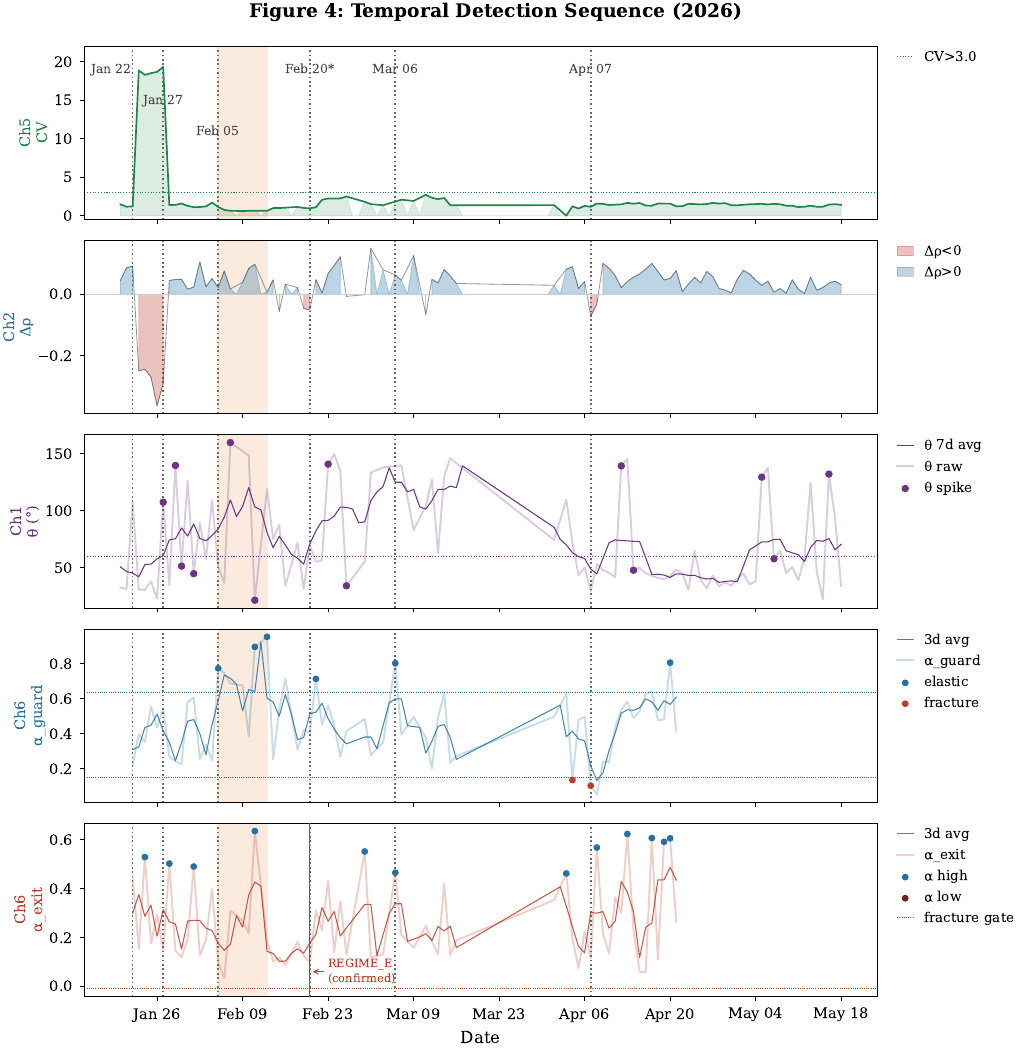}
  \caption{Selected temporal detection sequence across 67 observation
    windows (2026). Five panels share a common x-axis.
    \textbf{Ch5~CV}: thermodynamic precursor; spike Jan~23--26
    precedes the Jan~27 geometric response in the observed
    precursor sequence.
    \textbf{Ch2~$\Delta\rho$}: CCA agreement magnitude; sign
    transitions mark observer disagreement events.
    \textbf{Ch1~$\theta$}: CCA rotation angle; large swing
    events marked with dots.
    \textbf{Ch6~$\alpha_{\rm guard}$}: guard cluster soft
    alignment; dots mark large swings.
    \textbf{Ch6~$\alpha_{\rm exit}$}: exit cluster soft
    alignment; the REGIME\_E reference marker denotes the
    externally confirmed Feb~20 event window. Nearby
    $\alpha_{\rm exit}$ extrema reflect channel behavior
    and are not treated as independent causal attribution.
    Visual inspection suggests Ch2~$\Delta\rho$ transitions
    precede or coincide with Ch1~$\theta$ rotation events.
    Formal statistical validation is left for future work.}
  \label{fig:seq}
\end{figure*}

The six channels operate as a complementary system, not as
redundant sensors. A selected temporal subset of their
relationships is illustrated in Figure~\ref{fig:seq}.

In the observed event sequences, the firing order is
event-dependent. During the January 23--26 to January 27
sequence, Ch5 CV rose to 19.3 on January 26, followed by
Ch6 $\alpha_{\rm guard}$ shift on January 27. During the
February 20 event, Ch5 CV remained quiet (1.106) while
Ch2 $\Delta\rho$ and Ch1 $\theta$ detected observer
divergence alongside global EJT firing at $z = -4.38$.
The two observers do not follow a fixed firing order.
Their sequentiality is event-dependent and is itself
diagnostic: which observer fires first identifies what
kind of event is occurring.

Visual inspection of Figure~\ref{fig:seq} suggests a
consistent within-channel ordering: Ch2~$\Delta\rho$
transitions precede or coincide with Ch1~$\theta$ rotation
events, and $\theta$ rotation windows coincide with large
swings in Ch6~$\alpha_{\rm guard}$ and $\alpha_{\rm exit}$.
Formal statistical validation of this ordering is left
for future work.

Figure~\ref{fig:seq} illustrates the within-channel
temporal ordering observed across the 67-window dataset,
showing a selected channel sequence most relevant to the
observed temporal ordering. Channels Ch3~$\varepsilon$,
Ch4~$\sigma_z$, and $\dmg$ are omitted from this figure
for readability; they appear in the full forensic monitor
(Appendix~A).

The $\dmg$ discriminator: on February 05--13, large
$\alpha_{\rm guard}$ was accompanied by $\dmg = 5.17$,
supporting the forensic interpretation of a population
surge within the REGIME\_E pipeline label. On March 06,
an identical $\alpha_{\rm guard}$ signal was accompanied
by $\dmg = 2.55$, supporting the REGIME\_D interpretation
as internal reorganization rather than population flood.
Without $\dmg$, these events are difficult to distinguish
on $\alpha_{\rm guard}$ alone.

\subsection{Population Center Characterization}
\label{sec:pop_center}
Tor relay bandwidth exhibits moderate right-skew
(mean/median $= 1.4$) with isolated extreme outliers
(max/median $= 14\times$). These outliers caused sigmoid
saturation in the GRBM hidden layer at inference time,
motivating the $\Fvis$ bypass. For cluster center
computation, the arithmetic mean produces a ghost point
that no real relay occupies. The consensus-weight-weighted
median anchors the cluster center to the relay most
representative of actual Tor traffic load. This change
reduced exit cluster FPR from 26.3\% to 0.0\%.

\section{Event Taxonomy}

Based on the intersection of the six channels across
consecutive windows, the pipeline defines six geometric
event classifications requiring no causal attribution:
\textbf{PRECURSOR} (thermodynamic fragmentation before
geometric deformation), \textbf{REGIME\_S} (elastic
population surge absorbed without fracture),
\textbf{REGIME\_D} (localized geometric deformation),
\textbf{REGIME\_E} (stiff-axis fracture at the population
level), \textbf{REGIME\_K} (administrative maintenance
distinguished by forensic checklist), and \textbf{NORMAL}
(no gates fire).

The prefix REGIME indicates a geometric classification;
the suffix identifies the structural character of the
event. PRECURSOR and NORMAL are not prefixed REGIME
because they do not represent deformation events. Full
definitions, examples, and operational meanings follow.

Two of these classifications (REGIME\_S and REGIME\_K)
are defined theoretically but were not produced by the
pipeline classifier during the 67-window observation
period. Their definitions are retained because forensic
analysis identified their signatures within events the
classifier labeled REGIME\_E. All six classifications
are summarized in Table~\ref{tab:activations}.

\textbf{PRECURSOR.}
\emph{Definition}: Ch5 CV $> 3.0$ (GRBM thermodynamic
observer only). The CDAE geometric observer remains blind
to this event type because thermodynamic fragmentation
precedes any geometry-level deformation. This is the
earliest detectable signal in the pipeline.
\emph{Example}: January 23--26, 2026: CV reached 19.3,
the dataset maximum, across four consecutive windows.
All geometric channels remained near baseline.
January~22 was classified as REGIME\_E by the pipeline
(gate\_stiff fired) before the thermodynamic
fragmentation peak.
\emph{Meaning}: The relay population is fragmenting into
distinct behavioral groups in energy space, while the
geometric manifold remains near baseline. In the observed
January sequence, CV firing alone preceded a structural
transition in subsequent windows.

\textbf{REGIME\_S (Elastic Stretch).}
\emph{Definition}: Ch6 $\alpha_{\rm guard}$ z-score $> +2.0$,
cluster center shift magnitude exceeds the population median,
$\dmg$ exceeds threshold, sustained across three or more
consecutive windows.
\emph{Example}: February 05--13, 2026: $\dmg = 5.17$ during
the population surge from $\sim$10,500 to $\sim$19,640 relays.
$\alpha_{\rm guard}$ fired while $\alpha_{\rm exit}$ remained
quiet, and Ch5~CV was partially elevated but below the gate.
\emph{Meaning}: The network absorbed a massive relay influx
elastically, without structural fracture. The large $\dmg$
supports the interpretation that the surge consisted of
lightweight relays: the traffic-weighted cluster center
barely moved while the unweighted relay-count center
shifted substantially.
The guard cluster bent; it did not break.

\textbf{REGIME\_D (Deformation).}
\emph{Definition}: Ch6 $\alpha_{\rm guard}$ z-score $> +2.0$,
cluster center shift magnitude exceeds population median,
isolated to one or two windows. Ch5~CV remains below gate.
The CDAE geometric observer fires; the GRBM thermodynamic
observer remains quiet. This observer divergence
distinguishes REGIME\_D from PRECURSOR.
\emph{Example}: March 06, 2026: $\alpha_{\rm guard}$
z-score $= +3.36$, $\dmg = 2.55$, CV $= 1.85$. The
moderate $\dmg$ (compared to 5.17 during the
population-surge signature) supports interpreting this as
an internal guard cluster reorientation rather than a
population flood.
\emph{Meaning}: A localized coordinated shift in the
guard cluster geometry, visible only to the geometric
observer. No thermodynamic fragmentation and no population
surge; the event is structurally contained.

\textbf{REGIME\_E (Stiff Fracture).}
\emph{Definition}: Global EJT z-score $< -2.0$.
FPR $= 0.0\%$ on 24 stable windows. This is the only
gate with a verified 0.0\% FPR alongside Ch5~CV.
\emph{Example}: February 20, 2026: global EJT z-score
$= -4.38$ (4.38 standard deviations below the stable
baseline mean). Per-cluster $\alpha_{\rm exit}$ z-score
$= -1.342$ (sub-threshold), indicating the deformation
was distributed across the full population rather than
concentrated in the exit cluster. CCA also fired:
$\theta = 67.13^\circ$, $\Delta\rho = -0.0017$.
\emph{Meaning}: The full relay population mean moved
into the stiff load-bearing axes simultaneously. The
distributed nature of the signal (global fires, per-cluster
does not) is consistent with a network-wide routing
disruption rather than a targeted cluster-level event.
This is the dataset's only externally confirmed event.

\textbf{REGIME\_K (Administrative Maintenance).}
\emph{Definition}: Ch6 $\alpha_{\rm guard}$ z-score $< -2.0$,
cluster center shift magnitude greatly exceeding the
population median. A six-step forensic checklist is
required to distinguish coordinated administrative
maintenance from a hostile guard-layer attack.
\emph{Example}: April 07--08, 2026: $\alpha_{\rm guard}$
z-score $= -2.78$. Analysis of \texttt{days\_since\_restart}
revealed a strictly bimodal distribution: the returning
cohort had a mean restart age of 28 days, while the
non-returning cohort had a mean of 603 days. A ratio of
21:1 means the non-returning relays had been running
continuously 21 times longer. Under normal network churn,
relays restart for independent reasons at distributed
times; the restart-age distribution is unimodal and
diffuse. A bimodal distribution with a 21:1 age ratio
indicates two structurally distinct populations overlapping
in the same window: a coordinated fleet restart and
organic long-running relay retirement.
\emph{Meaning}: A coordinated administrative fleet restart
produces a stiff-axis collision in the guard cluster
because a large cohort of relays departs and is rapidly
replaced. High consensus weight on Day-1 arriving relays
indicates trusted infrastructure returning, not a hostile
takeover.

\textbf{NORMAL.}
\emph{Definition}: No detection gates fire.
\emph{Example}: The majority of the 67 observation windows.
\emph{Meaning}: Network stability. Population movements
are absorbed elastically through the soft subspace.

\section{Results}

\textbf{The Proven Event: February 20, 2026.} On February 20, 2026, the pipeline registered a significant
structural anomaly classified as REGIME\_E. This event
aligns temporally with a confirmed Cloudflare BYOIP
configuration event resulting in BGP prefix withdrawals
(4 full withdrawals confirmed by RIPE NCC Routing Information Service (RIPE RIS) routing data~\cite{ripe-ris} at
approximately 16:00~UTC). BGP, the Border Gateway Protocol,
determines how traffic flows between internet providers.
A BGP prefix withdrawal means a block of IP addresses
became unreachable from the broader internet.

Two independent signals fired. First, Ch2/Ch1 CCA detected
significant observer divergence: $\theta = 67.13^\circ$
and $\Delta\rho = -0.0017$, indicating the shared
structural axis rotated sharply and agreement weakened.
Second, the global EJT registered z-score $= -4.38$,
meaning the full relay population mean moved 4.38 standard
deviations into the stiff load-bearing axes. The per-cluster
$\alpha_{\rm exit}$ z-score was $-1.342$, below the
$-2.0$ gate threshold: the deformation was distributed
across the population rather than concentrated in the
exit cluster.

Forensic analysis examined whether this event was caused
by relay departures from Cloudflare infrastructure. By
comparing the Tor consensus at February 19 23:00~UTC with
February 20 23:00~UTC, we identified 116 relays present
on February 19 but absent on February 20. We
cross-referenced the IP addresses of these 116 departed
relays against 11 confirmed Cloudflare IPv4 prefixes from
cached RIPE RIS BGP data. \emph{Zero of 116 departed relay
IPs matched any Cloudflare prefix. The relay departure
hypothesis is formally falsified.}

The observed geometric anomaly is therefore consistent with
connectivity degradation: the Cloudflare BGP withdrawal could
have affected reachability paths while no Tor relay actually
left the consensus. The network's effective connectivity
changed without a topology-level relay-departure signature.
Relay-level geometry captures effective network state rather
than explicit topology.

\textbf{Feature Attribution.} The encoder's stiff subspace concentrates on geographic
and temporal relay features: latitude, longitude, and
\texttt{days\_since\_restart}. This was confirmed by a
per-cluster Jacobian analysis conducted across all role
clusters, which evaluated the encoder sensitivity at each
role cluster center and
identified which of the 17 input features contributed
most to the stiff eigenvectors. Geographic and temporal
features dominated the load-bearing directions. The global EJT signal at
z-score $= -4.38$ reflects population-wide mean
displacement along these directions.

The 116 true consensus departures were \emph{middle-heavy}
(51.7\% Middle-role relays) with exits comprising only
20.7\%. In the Tor network, Middle relays route traffic
internally without touching either circuit endpoint.
A departure cohort described as middle-heavy means the
majority of departing relays were Middle-role, ruling out
a targeted exit-layer attack. The signal is consistent
with broad routing disruption affecting all relay types.

\textbf{Event-Gate Summary.} The event-gate activation pattern is summarized in
Table~\ref{tab:activations}. No single channel captures all
events: the combination of channels is the detection
framework.

\textbf{The Thermodynamic Precursor Sequence.} In the two weeks preceding the January 27 REGIME\_D
reorganization, Ch5 CV reached its dataset maximum of 19.3
on January 26, indicating maximum thermodynamic
fragmentation before any structural deformation was visible
to the CDAE geometric observer. The sequence (PRECURSOR
on January 23--26, REGIME\_D on January 27) is one
observed instance of thermodynamic signal preceding
geometric deformation. Whether this sequentiality is
consistent across event types is an open question.

\textbf{False Positive Analysis and Administrative Behavior.} All gate activations across 67 windows are cataloged in
Table~\ref{tab:activations}. Events lacking external
attribution are classified strictly by geometric signature
and are not claimed as confirmed attacks.

The April 08 event demonstrates the pipeline's ability to
differentiate administrative maintenance from hostile
structural degradation. The returning cohort had a mean
restart age of 28 days vs.\ 603 days for the non-returning
cohort, a ratio of 21:1. Two populations this different in
operational age cannot be explained by normal network churn.
The REGIME\_K forensic checklist confirmed the fleet
restart interpretation.

\textbf{Coordinated Relay Deployment: February 05--13.} Forensic analysis of the February 05--13 population surge
identified a dominant operator pattern.
The contact identity \texttt{printerexpert@mail2tor.com}
accounted for 7,600 unique relay fingerprints and 227,929
descriptors (55\% of 414,443 total surge descriptors).
Four independent uniformity indicators show zero variance
across all 227,929 descriptor records: identical platform string,
identical bandwidth configuration (12~MB/s cap with
mean $=$ median $=$ max), identical burst configuration,
and 96.8\% exit-configured policy. No collection of
independent operators producing 227,929 descriptor records
across 7,600 relay fingerprints would produce this uniformity
by coincidence. Independent
operators typically deploy different Tor versions,
configure bandwidth limits according to their own
available connection capacity, and choose exit policies
based on individual preferences and legal constraints.
The result is a diverse distribution across all four
indicators. Zero variance simultaneously across all four
is the signature of a single configuration template
applied at scale.

IP addresses recovered from the Tor consensus archive for
2,058 of the 7,600 unique fingerprints are 100\%
residential broadband, concentrated in the USA (43.4\%)
and Argentina (34.3\%). No Cloudflare address space was
detected across the 2,056 checkable IPs compared against
26 Cloudflare prefix ranges. The Tor directory authorities rejected 72.9\%
of the deployment. Operator intent is unknown; full
investigation is the subject of a subsequent paper.

\clearpage
\onecolumn
\begin{table}[p]
  \vspace{-0.7em}
  \centering
  \captionof{table}{Gate Activation Log (67-Window Observation Period).
    Feb~20 is the only externally confirmed event.
    Apr~03 values from sweep CSV:
    $\theta = 109.74^\circ$, EJT z-score $= -0.71$
    (sub-threshold), CV $= 0.008$, event type MODE\_F.
    Dec~20 2025 predates the sweep window.
    MODE\_F denotes windows where gate\_geometric fired but
    the global EJT z-score remained sub-threshold ($z > -2.0$),
    indicating a geometric reorientation without confirmed
    stiff-axis fracture. MODE\_F events fall outside the
    six-class taxonomy and are reported as observed pipeline
    output. Events marked Uninvestigated have no external
    attribution and no forensic experiment was conducted;
    they are documented as observed gate activations
    establishing the operational noise floor.}
  \label{tab:activations}
  \small
  \setlength{\tabcolsep}{6pt}
  \renewcommand{\arraystretch}{2.2}
  \begin{tabularx}{\textwidth}{@{}p{1.5cm}p{2.2cm}p{2.8cm}Xp{2.1cm}@{}}
    \toprule
    \textbf{Date} & \textbf{Channels} & \textbf{Values}
    & \textbf{Classification and Finding} & \textbf{Status} \\
    \midrule
    Jan 23--26 & Ch5 CV
      & CV $> 3.0$ (max 19.3)
      & PRECURSOR: thermodynamic fragmentation preceding
        Jan~27 transition. All geometric channels quiet.
        Jan~22 classified as REGIME\_E by pipeline
        (gate\_stiff fired).
      & Investigated \\
    Jan 27 & Ch6 guard elastic
      & z-guard $> +2.0$
      & REGIME\_D: internal reorganization. CDAE fires,
        GRBM quiet. Observer divergence confirmed.
      & Investigated \\
    Feb 05--13 & Ch6 guard + $\dmg$
      & z-guard $> +2.0$, $\dmg = 5.17$
      & REGIME\_E (pipeline label). Coordinated residential
        relay surge. 7,600 single-operator relays. 72.9\%
        rejected by Tor directory authorities. Distinguished
        from Feb~20 REGIME\_E by $\dmg = 5.17$ (population
        surge) vs Feb~20 $\dmg \approx 2.88$ (connectivity
        degradation).
      & Investigated \\
    \rowcolor{gray!15}
    Feb 20 & Ch2 $\Delta\rho$, Ch1 $\theta$, Global EJT
      & $z = -4.38$, $\theta = 67.13^\circ$,
        $\Delta\rho = -0.0017$
      & REGIME\_E: Cloudflare BGP withdrawal confirmed
        (RIPE RIS). Relay departure hypothesis falsified:
        0/116 departed relay IPs matched Cloudflare
        prefixes. Connectivity-degradation interpretation
        supported.
      & \textbf{CONFIRMED} \\
    Mar 06 & Ch6 guard elastic
      & z-guard $= +3.36$, CV $= 1.85$
      & REGIME\_D: CDAE fires, GRBM quiet. Localized guard
        expansion invisible to thermodynamic observer.
      & Investigated \\
    Apr 03 & Ch6 guard elastic
      & $\theta = 109.7^\circ$, z $= -0.71$ (MODE\_F)
      & CCA rotation elevated but EJT z-score
        sub-threshold. Cause uninvestigated.
      & Uninvestigated \\
    Apr 07--08 & Ch6 stiff fracture
      & z-guard $= -2.78$
      & REGIME\_E (pipeline label). Forensic checklist
        confirms administrative fleet restart, not hostile
        event. Returners mean 28 days vs.\ non-returners
        603 days (21:1). REGIME\_K is the forensic
        interpretation; REGIME\_E is the geometric
        classification.
      & Investigated \\
    \bottomrule
  \end{tabularx}
  \vspace{-0.7em}
\end{table}
\clearpage
\twocolumn


\begin{figure*}[t]
  \centering
  \includegraphics[width=\textwidth]{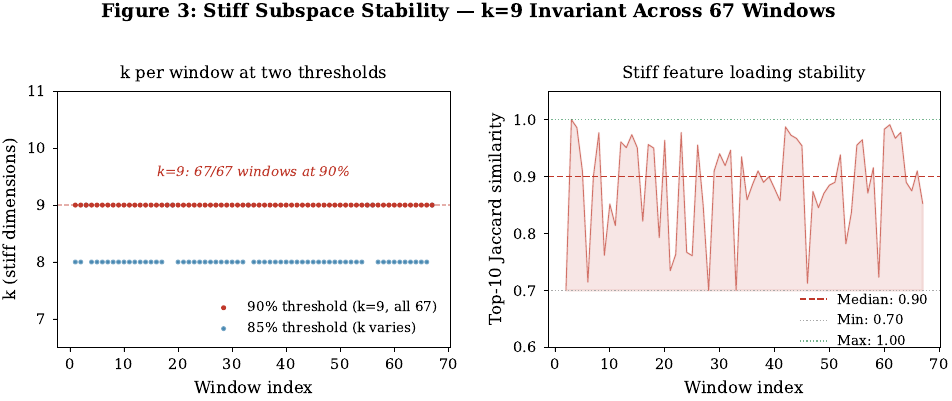}
  \caption{Stiff subspace stability. Across 67 observation windows,
    $k=9$ remains invariant at the 90\% trace-mass threshold, while
    Top-10 Jaccard similarity shows stable stiff feature loadings
    across consecutive windows.}
  \label{fig:k9}
\end{figure*}

\clearpage

\section{Limitations}
\label{sec:limitations}

The primary limitation is the sample size of ground truth.
We present one externally confirmed event (February 20
Cloudflare withdrawal). This is a fundamental observational
constraint of the domain: Tor withholds packet visibility,
flow telemetry, and path truth by design. Under these
constraints, one externally anchored reference event is a
substantial calibration point rather than a weakness.

The Ch1 CCA rotation gate ($\theta$) exhibited 49\% FPR at
the $60^\circ$ threshold and is invalid as a standalone
alarm. Channels Ch3 ($\varepsilon$) and Ch4 ($\sigma_z$)
are computed as proxies; the VAE variance head was retired
due to NaN instability and the model operates as a
deterministic autoencoder. These channels provide
supporting signal only.

CCA identifies linear shared structure only. The CDAE
contraction penalty enforces local linearity in the stiff
subspace, partially satisfying this assumption. Nonlinear
inter-observer structure, particularly in the GRBM hidden
activations, is not captured. Subtle events perturbing only
nonlinear observer agreement may evade detection.

The framework relies on Tor-specific domain knowledge.
Guard, Middle, and Exit role clusters are defined by the
Tor protocol. Generalizing to other networks requires
mapping equivalent structural roles. The constraint space
(17 clean features, $\lambda_c = 0.001$, 32 latent
dimensions) is chosen rather than imposed; different
choices produce different geometry.

Geometric anomalies do not equate to causal attribution.
The pipeline detects what changed geometrically, not why.
The gap between a confirmed structural anomaly and an
attributed attack vector requires external telemetry to
bridge.

The CDAE encoder is frozen at the January 19 to
February 09, 2026 training window. Periodic retraining or
continual learning approaches are required for sustained
operational validity.

\section{Future Work}

\textbf{Intra-observer rotational dynamics.}
CCA measures inter-observer agreement between the two
observer outputs. The missing layer may be intra-observer
rotational dynamics: how the energy distribution rotates
inside the GRBM space, and how the stiff eigenvectors
rotate inside the CDAE space, independently. Comparing
these internal rotation signatures may reveal a quantified
thermodynamic-to-geometric propagation lag. The January~22
to February~20 window (28 days) is the ideal test case.
This is the central research question of a subsequent paper.

\textbf{Mechanistic theory.}
Why $k = 9$ emerges and why certain features dominate the
stiff subspace remains open. The encoder independence test
(five independent CDAEs with identical constraints) and
feature ablation experiments are the next steps toward
mechanistic closure.

\textbf{Fixed-reference CCA.}
Per-window refit produces a $\theta$ signal lacking temporal
coherence for accumulation. A fixed-reference CCA would
enable $\theta_{\rm acc}$, a rolling sum of accumulated
excess observer disagreement, potentially constituting a
second geometric precursor channel complementary to Ch5~CV.

\textbf{Kernel CCA.}
Standard CCA identifies only linear shared structure.
Kernel CCA would capture nonlinear inter-observer agreement,
potentially revealing subtle events that linear CCA misses.

\textbf{Feature ablation and second-order engineering.}
The stiff subspace is dominated by first-order geographic
and temporal features. These may be loud rather than
meaningful. Second-order features such as trust
accumulation rate and bandwidth efficiency may carry richer
structural information.

\textbf{The null geodesic.}
A sophisticated actor aware of the stiff axis can operate
through soft directions without triggering geometric gates.
The February 05--13 coordinated deployment, if confirmed
as adversarial, would be the first empirical evidence of
this threat model.

\textbf{Cross-network generalization.}
The Tor relay population is one instance of a large
behavioral population: thousands of nodes, each producing
daily feature snapshots, with no ground-truth labels.
BGP routing tables, DNS resolver populations, and
enterprise endpoint fleets share the same structure.
Applying the identical pipeline (frozen encoder, EJT
decomposition, dual-observer CCA) to these populations
would test whether a stable stiff subspace emerges
naturally from their behavioral geometry, as it did in
Tor. If $k$ is invariant there too, the framework
generalizes. If it is not, the invariance is
Tor-specific and the claim must be narrowed.

\textbf{Real telemetry and attribution closure.}
On networks with Zeek connection logs, NetFlow, or BGP
event streams, the geometric signal maps directly to
flow-level causal evidence. The geometry provides early
detection. The telemetry closes attribution.

\section{Conclusion}

This work presents a geometric anomaly detection framework
for large behavioral populations, validated on the Tor
anonymity network across 67 consecutive daily observation
windows. The framework requires no labeled training data,
no attack signatures, and no manual threshold tuning beyond
a statistically derived baseline.

The central finding is that the encoder's stiff subspace,
the nine dimensions capturing the load-bearing structure
of the relay population, is invariant across the full
observation period. This invariance was not chosen. It
emerged from the intersection of the training data and the
architectural constraints. The geometry belongs to the
data, not to the design. Structure precedes geometry.

Two complementary observers detect structural deformation
through different mechanisms. In the observed event
sequences, the firing order is event-dependent: during the
January 23--26 sequence the GRBM thermodynamic observer
fired first; during the February 20 event the geometric
observers fired while the GRBM remained quiet. When they
disagree, the nature of their disagreement identifies
the event type.

One event was externally confirmed. The February 20, 2026
Cloudflare BYOIP configuration event aligned with a global
EJT signal of $z = -4.38$ within a pipeline whose structural
agreement was validated at $16.8\sigma$ above the Monte Carlo
null. Forensic analysis formally falsified the relay departure
hypothesis: zero of 116 true consensus departures matched
any Cloudflare address prefix.
The pipeline detected connectivity degradation without
topology change, a failure mode not visible to standard
relay-count monitoring.

A secondary forensic investigation identified a coordinated
mass deployment of 7,600 relays from residential IP
infrastructure during the February 05--13 population surge.
Tor's directory authorities rejected 72.9\% of the deployment;
operator intent remains unknown and is reserved for a
subsequent paper. The remaining windows are geometric
classifications only: their causes are unknown, and they
establish the operational noise floor. This is the boundary of
the claim: the paper measures deformation, validates one
reference event, and leaves unattributed regimes open.

\clearpage
\appendix
\renewcommand{\thefigure}{A\arabic{figure}}
\setcounter{figure}{0}
\section{Full Forensic Monitor}

Complete forensic monitor for the 67-window observation
period, serving as the primary audit record. All
detection channels, gate activations, and event
classifications are shown simultaneously. Readers
wishing to verify any specific window's signal values
should refer to this figure alongside Table~1.

\begin{figure*}[p]
  \centering
  \includegraphics[width=\textwidth,height=0.92\textheight,keepaspectratio]{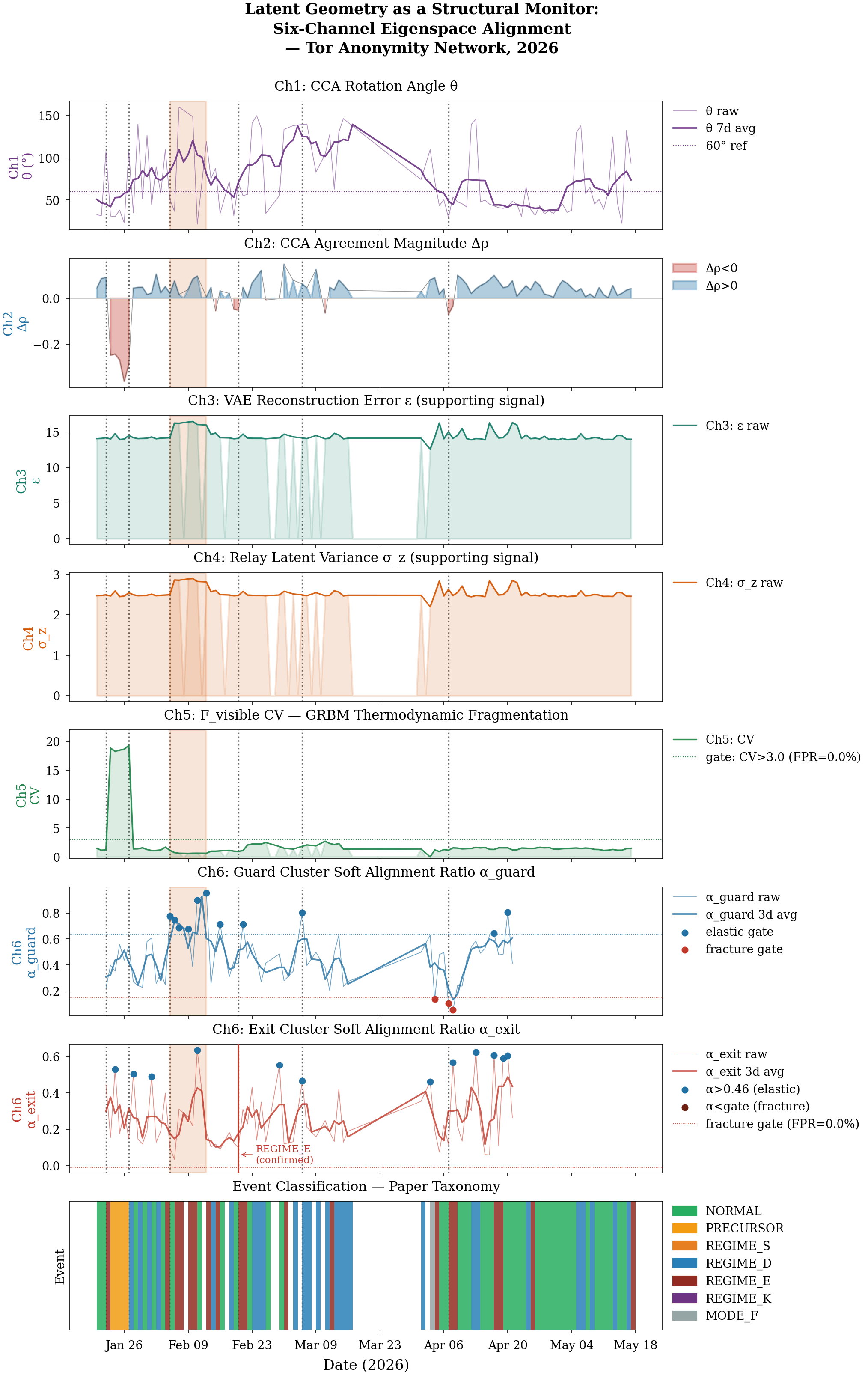}
  \caption{Complete forensic monitor for the 67-window observation period,
    serving as the primary audit record. All detection channels, gate
    activations, and event classifications are shown simultaneously.
    Readers wishing to verify any specific window's signal values should
    refer to this figure alongside Table~1.
    A data collection gap is visible across all panels between
    approximately March~17 and March~31, 2026, corresponding
    to 15 missing daily relay snapshots in the ingestion
    pipeline. No data was collected for those dates; the gap
    is not interpolated and reflects a genuine absence in the
    source data. All detection results reported in this paper
    are unaffected by this gap as the confirmed events fall
    outside this window.}
  \label{fig:dashboard}
\end{figure*}


\end{document}